\documentclass[sigconf,10pt]{acmart}
\usepackage{graphicx}
\usepackage{booktabs}

\AtBeginDocument{%
  \providecommand\BibTeX{{%
    \normalfont B\kern-0.5em{\scshape i\kern-0.25em b}\kern-0.8em\TeX}}}

\setcopyright{none}

\acmConference[MobiSys '26]{The 24th ACM International Conference on Mobile Systems, Applications, and Services}{June 21--25,
  2026}{Cambridge, UK}

\renewcommand\footnotetextcopyrightpermission[1]{}

\graphicspath{{./images/}}
\begin{document}

\title{Poster: EdgeCitadel --- Hybrid NATS--MQTT Orchestration for Edge Multi-Agent Systems}

\author{Zhonghao Zhan}
\affiliation{%
  \institution{Imperial College London}
  \city{London}
  \country{United Kingdom}}
\email{zzhan@ic.ac.uk}

\author{Yefan Zhang}
\affiliation{%
  \institution{Independent Researcher}
  \city{Seattle}
  \country{United States}}
\email{zhangyefan752@gmail.com}

\author{Hamed Haddadi}
\affiliation{%
  \institution{Imperial College London}
  \city{London}
  \country{United Kingdom}}
\email{h.haddadi@imperial.ac.uk}

\begin{abstract}
Edge-resident AI agents increasingly span home servers, IoT hubs, laptops, and phones, yet their coordination stacks still assume cloud-style transports or a central relay.
We present \textsc{EdgeCitadel}\footnote{Code and artifact: \url{https://github.com/zhonghaozhan/EdgeCitadel}}, an edge multi-agent orchestration platform built around a single NATS~2.10 server with the built-in MQTT adapter.
The design combines MQTT connectivity for heterogeneous agents, JetStream-backed persistence and replay for backend services, direct peer delegation over a shared subject namespace, and a passive aggregator that visualizes and stores traffic without sitting on the delivery path.
Our poster highlights the migration from MQTT relay prototypes (common in IoT communication) to the current hybrid architecture and demonstrates a working cross-device testbed spanning ARM64, x64, and Android clients.
\end{abstract}

\maketitle

\section{Motivation}

AI agents are moving from cloud sandboxes to persistent edge deployments on heterogeneous devices~\cite{satyanarayanan-edge}.
Existing frameworks such as AutoGen~\cite{autogen}, LangGraph~\cite{langgraph}, and the OpenAI Agents SDK~\cite{openai-agents} provide useful orchestration abstractions, but they do not directly solve deployment-time orchestration across mixed edge hardware. Anthropic notes that cross-session coordination requires explicit external state~\cite{anthropic-harness}; IoT conflict surveys confirm that co-located applications need structured arbitration~\cite{huang-conflicts}. Moreover, agents on edge do not choose their message infrastructure; instead, they inherit MQTT-style pub/sub as the de facto coordination layer.

\textsc{EdgeCitadel} targets this gap.
Compared with MQTT-only relay prototypes, the current system makes three deliberate design moves:
(1)~a single NATS backbone with the built-in MQTT adapter~\cite{nats},
(2)~direct agent-to-agent delegation over structured subjects instead of backend-mediated relaying, and
(3)~a passive observer that records and visualizes system activity without becoming part of the delivery path.

\section{System Architecture}

\begin{figure}[t]
\centering
\includegraphics[width=\columnwidth]{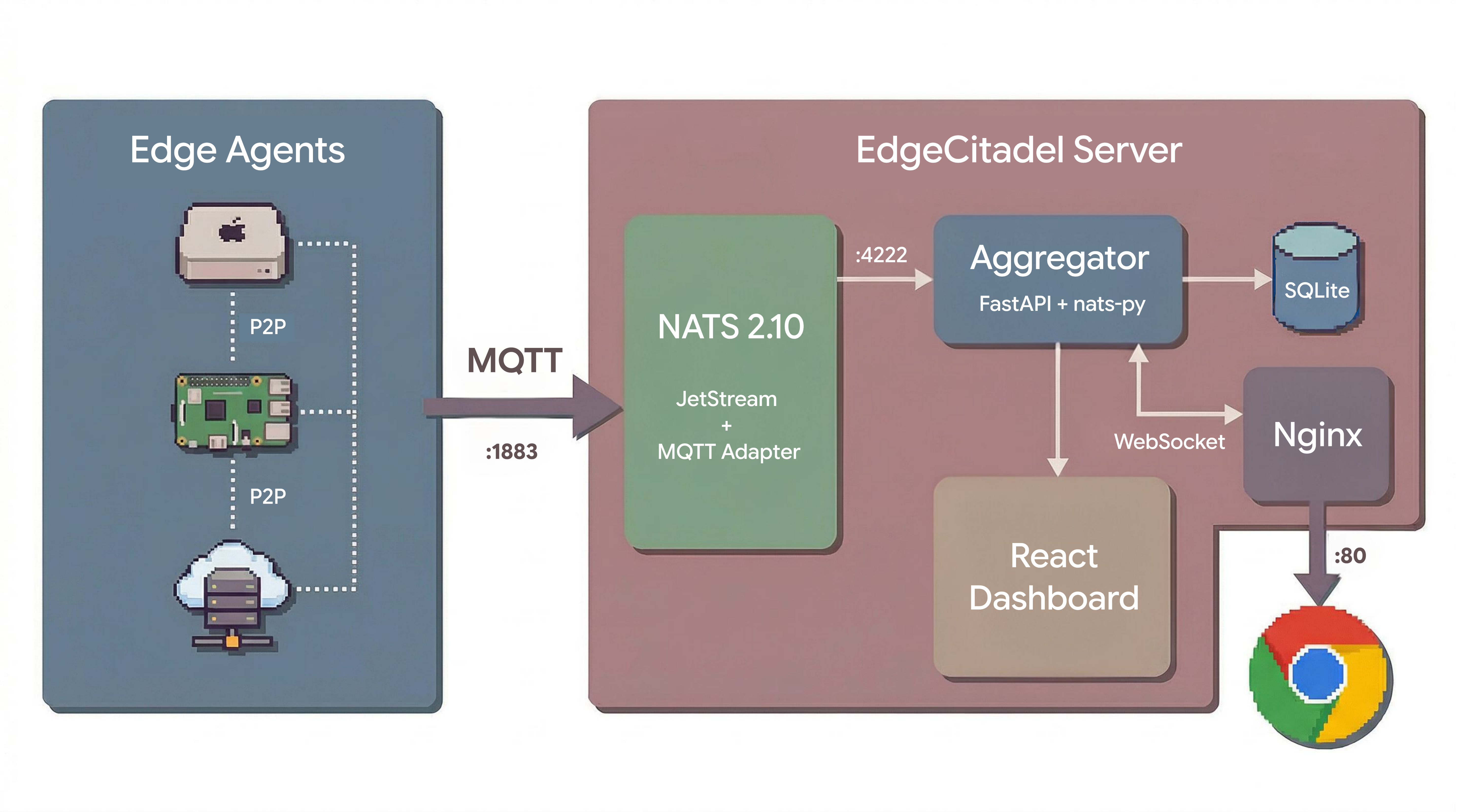}
\caption{EdgeCitadel architecture. Edge agents connect via MQTT, backend services use native NATS, and a passive aggregator persists and visualizes traffic.}
\label{fig:arch}
\end{figure}

As shown in Figure~\ref{fig:arch}, \textsc{EdgeCitadel} uses one NATS~2.10 server as its messaging backbone.
Edge agents connect via MQTT on port~1883; the aggregator and other backend services use the native NATS protocol on port~4222.
NATS automatically translates between the two (e.g., MQTT inbox $\leftrightarrow$ NATS inbox), so there is no separate MQTT broker and no split authentication.
Nginx reverse-proxies the dashboard UI, the REST API, and WebSocket streams to the browser.

\paragraph{Hybrid transport and persistence.}
The hierarchy is built around \texttt{agents.\{id\}.\{action\}}, \texttt{tasks.\{id\}.\{phase\}}, and \texttt{system.broadcast}.
JetStream saves the corresponding message families into a durable conversation stream for future audits, while a key-value bucket (\texttt{AGENT\_STATE}) maintains live agent metadata for independent peer discovery.

\paragraph{Passive observation.}
The FastAPI aggregator subscribes to wildcard NATS subjects (\texttt{agents.>}, \texttt{tasks.>}, \texttt{system.>}), stores every message in SQLite, and streams structured events to the React dashboard over WebSocket.
Dashboard-initiated commands are published to NATS by the aggregator, but all agent-to-agent traffic flows directly through the broker without aggregator involvement, including P2P delegation.
This means the aggregator can crash or fall behind without disrupting inter-agent coordination.

\paragraph{Delegation with guardrails.}
Agents can issue direct peer delegations by writing to a target inbox subject and linking replies with correlation IDs.
Three mechanisms keep delegation bounded in case of indirect prompt injection in agent pipelines~\cite{greshake-injection}: a content-hash computed over the first 200 characters of each payload to detect A$\to$B$\to$A cycles, a per-hop \texttt{chain\_depth} counter hard-limited to~5, and a 90-second timeout with at most 5 concurrent outstanding requests.
After delegations return, the origin agent's LLM summarizes and analyzes results and may initiate further rounds up to a configurable depth limit.

\section{Poster Demonstration}

\paragraph{Testbed.}
We deploy on three classes of commodity hardware: an M4 Mac~Mini (ARM64, 16\,GB), an Intel NUC11 (x64, 4~cores, 8\,GB, Ubuntu~24.04), and a Pixel~4 (Android~14), connected over a Tailscale mesh VPN.
Agents are hosted using OpenClaw~\cite{openclaw} as the local LLM/Agent runtime; physical IoT devices (Philips Hue lights, JBL speakers, Reolink camera, LG smart TV) are controlled through Home Assistant and Android UI automation adapters.

\begin{figure}[t]
\centering
\includegraphics[width=\columnwidth]{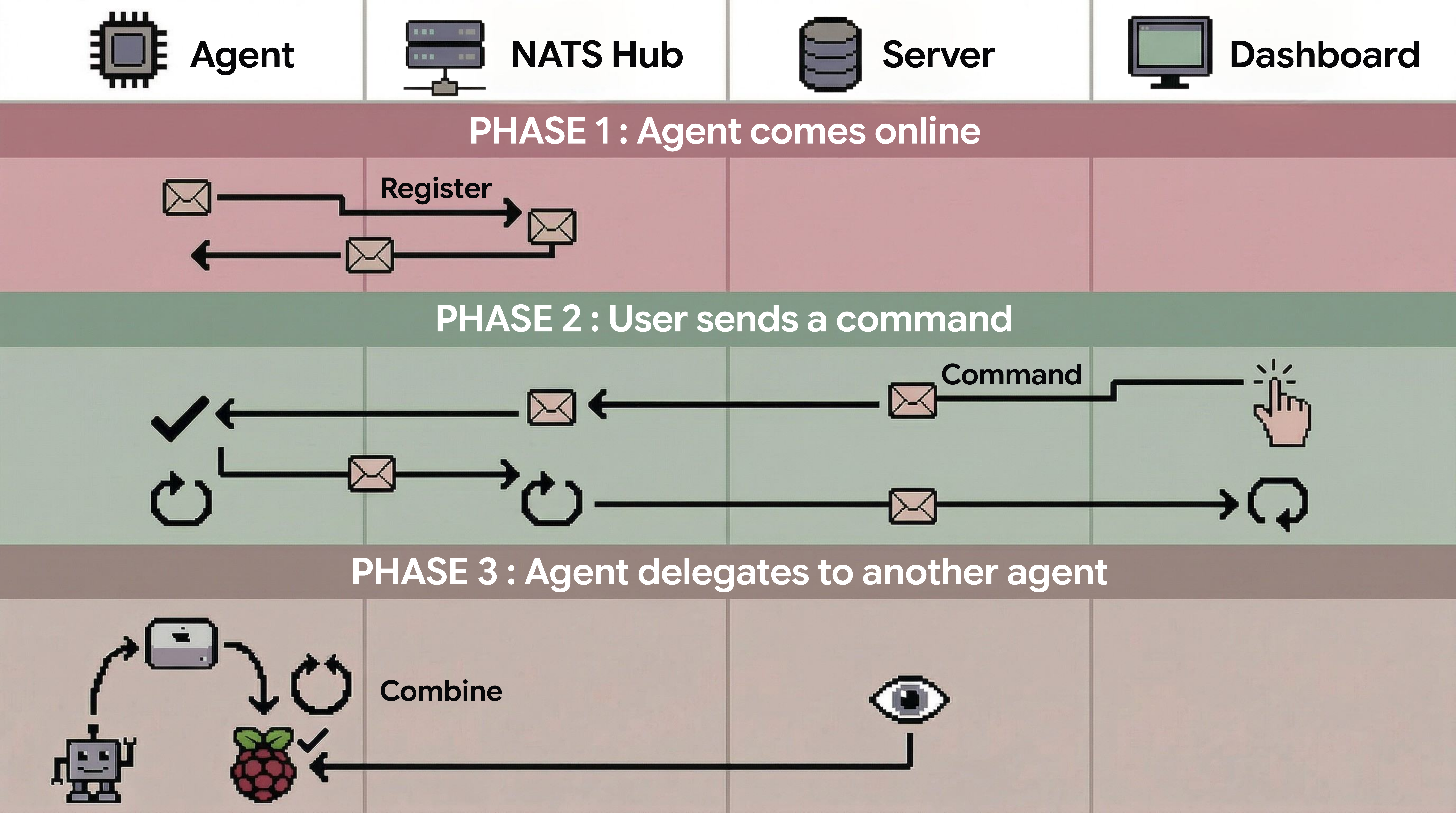}
\caption{Communication flow through three phases. (1)~An agent comes online: it registers with the NATS hub, the server forwards the event. (2)~A user sends a command via the dashboard; it is routed through the server and hub to the agent, whose response follows the reverse path. (3)~An agent delegates to a peer directly through the hub; the server monitors all traffic.}
\label{fig:flow}
\end{figure}

\paragraph{Scenarios.}
The poster demonstrates three scenarios that mirror the communication phases in Figure~\ref{fig:flow}:\newline
(1)~a dashboard command that triggers coordinated multi-device actuation, walking through the full command path:\newline
dashboard$\to$server$\to$hub$\to$agents$\to$response; \newline
(2)~a cross-domain query in which one agent autonomously delegates to a peer through the hub, with the server capturing the entire chain passively; and
(3)~a disconnect/rejoin sequence in which a killed agent re-registers over MQTT and the JetStream stream replays missed messages, with the dashboard reflecting the status transition in real time.

\paragraph{Preliminary evidence.}
Table~\ref{tab:results} summarizes orchestration outcomes from the testbed.
End-to-end latency is dominated by LLM inference at each agent hop (1–5 s), not by messaging infrastructure; MQTT-to-NATS round-trip overhead is under 4 ms at p95, and P2P delegation adds under 5 ms per hop. The passive aggregator persisted all 50 test messages without loss, and re-registration after an offline event completes in under 17 ms (p95).

\begin{table}[t]
\caption{Infrastructure overhead measured on a single-host Docker Compose deployment (20 iterations per metric). We report containerized results as a reproducible baseline; physical multi-host measurements will vary with network topology and hardware.}
\label{tab:results}
\small
\begin{tabular}{@{}lr@{}}
\toprule
\textbf{Metric} & \textbf{Result} \\
\midrule
\multicolumn{2}{@{}l}{\emph{Hybrid Transport}} \\
\quad MQTT$\leftrightarrow$NATS round-trip (p95) & 3.51\,ms \\
\quad Message persist latency (p95) & 19.1\,ms \\
\midrule
\multicolumn{2}{@{}l}{\emph{P2P Delegation}} \\
\quad Single-hop round-trip (median) & 2.78\,ms \\
\quad Single-hop round-trip (p95) & 4.20\,ms \\
\midrule
\multicolumn{2}{@{}l}{\emph{Passive Observation}} \\
\quad Message persistence completeness & 50/50 \\
\quad P2P delivery without aggregator & Yes \\
\quad Dashboard event latency (p95) & $<$19\,ms \\
\midrule
\multicolumn{2}{@{}l}{\emph{Agent Lifecycle}} \\
\quad Re-registration after offline (median) & 12.9\,ms \\
\quad Re-registration after offline (p95) & 16.5\,ms \\
\quad Heartbeat status detection & Correct \\
\bottomrule
\end{tabular}
\end{table}

\section{Discussion and Artifact}

The system is packaged as a reproducible Docker Compose deployment with a one-line quick start script for provisioning agents, and end-to-end Playwright tests covering registration, heartbeat monitoring, command pipelines, delegation chains, and task lifecycle transitions.

Our main takeaway from replacing a naive MQTT relay with the current architecture is that edge multi-agent systems benefit from treating the messaging layer as first-class infrastructure rather than hidden middleware.
Using NATS with the MQTT adapter preserves compatibility with lightweight IoT clients while giving backend services access to durable streams, structured subjects, and auditable traces. These are capabilities that previously required substantial custom engineering on a plain MQTT relay used in most IoT deployments. We hope the poster sheds light on the future development of agent runtimes regarding the practice of orchestrating edge multi-agent systems.

\bibliographystyle{ACM-Reference-Format}
\bibliography{references}

\end{document}